\documentclass[journal,10pt]{IEEEtran}

\usepackage{cite}
\usepackage{graphicx}
\usepackage[cmex10]{amsmath}
\usepackage{amsthm}
\usepackage{amsfonts}
\usepackage{caption}
\usepackage{epstopdf}
\usepackage{amssymb}
\usepackage{float}
\usepackage[export]{adjustbox}

\usepackage{color,soul}
\usepackage{subfig}
\usepackage{csquotes}
\usepackage{enumitem}
\usepackage{adjustbox,lipsum}
\usepackage{tabularx,booktabs}
\usepackage{array}
\newcolumntype{L}[1]{>{\raggedright\let\newline\\\arraybackslash\hspace{0pt}}m{#1}}
\newcolumntype{C}[1]{>{\centering\let\newline\\\arraybackslash\hspace{0pt}}m{#1}}
\newcolumntype{R}[1]{>{\raggedleft\let\newline\\\arraybackslash\hspace{0pt}}m{#1}}

\newcommand\numberthis{\addtocounter{equation}{1}\tag{\theequation}}
\newcounter{mytempeqncnt}
\newcounter{mytempeqncnt2}

\begin{document}
\title{Analytical Framework of Beamwidth Selection for RT-ICM Millimeter-Wave Clusters}
\author{Yavuz Yaman, \IEEEmembership{Student Member, IEEE}, and Predrag Spasojevic, \IEEEmembership{Senior Member, IEEE}
\thanks{The authors are with Department of Electrical and Computer Engineering, Rutgers University, Piscataway, NJ 08854 USA (e-mail:yavuz.yaman@rutgers.edu; spasojev@winlab.rutgers.edu).}}

\maketitle

\begin{abstract}
\boldmath
Beamforming for millimeter-wave (mmWave) communications is well-studied in the physical layer (PHY) based on the channel parameters to develop optimum receiver processing techniques. However, even before signal processing, antenna structure and radiation parameters affect the beamforming performance primarily. For example, in contrast to common belief, narrow beamwidth may result in degraded beamforming performance. In order to address the impairments such as beam misalignments, outage loss, tracking inability, blockage, etc., an optimum value of the beamwidth must be determined. In this paper, assuming a communication system that creates a beam per cluster, we theoretically investigate the beamwidth and received power relation in the cluster level mmWave channels. We adopt uniform linear array (ULA) antenna structure and formulate its antenna gain with respect to the beamwidth. Two beam models are considered for the main lobe of the array pattern, rectangular and triangular, to approximate the best and worst scenarios, respectively. For the channel, we derive beamwidth-dependent extracted power expressions for two intra-cluster channel models, IEEE 802.11ad and our previous work based on ray-tracing (RT-ICM). Combining antenna and channel gains, in case of a beam misalignment, we find that the optimum beamwidth that maximizes the received power is larger than the alignment error when the error itself is larger than the standard deviation of the cluster power-angle spectrum. Once the alignment error is smaller than the standard deviation, we confirm that the optimum beamwidth converges zero. Performing asymptotic analysis of the received power, we give the formulation and insights that the practical nonzero beamwidth values can be achieved although sacrificing subtle from the maximum received power. Our analysis shows that to reach $\%95$ of the maximum power for an indoor mmWave cluster, a practical beamwidth of $7^{\circ}-10^{\circ}$ is enough, which can be created with $18-20$ antenna elements. In the simulation section, we show that the expressions given by the analysis match to the simulated results.
\end{abstract}
\begin{IEEEkeywords}
\emph{millimeter wave, beamforming, intra-cluster, 60 GHz, 28 GHz, spatial filtering, power angle profile, antenna arrays, beamwidth}
\end{IEEEkeywords}

\section{Introduction}\label{intro}

\IEEEPARstart{M}{illimeter}-wave (mmWave) communication has several advantages over the current wireless bands such as higher throughput, lower latency, reduced interference, and increasing network coordination ability. Nevertheless, high path loss is the significant drawback of mmWave channels. To overcome, beamforming is proposed as a substantial solution with the availability of large array usage in a small-scale area. On the other hand, due to the sparse nature of mmWave channels, clusters are generally spatially-separated \cite{Maltsev_conference}. That further allows creating a beam for each cluster, both in the transmitter and the receiver end which, in turn, yields increased performance in multi-input-multi-output (MIMO) and massive MIMO applications \cite{11ay}. Significant contribution is published for the receiver processing aspect of beamforming, including optimum transmitter and receiver design \cite{Ayach} with array antennas and beamforming protocols \cite{Yaman}. However, maximizing the beamforming efficiency can be challenging due to the misalignments, weak tracking ability, blockages, outage loss, etc. which requires channel knowledge in the angular domain. Specifically, beams with non-optimized beamwidths may increase inter-beam interference and wasted energy, i.e. outage rate, or even cause a link failure easily when combined with beam misalignment. As a result, while the requirement of the accuracy on the beam alignment to the cluster angle of arrival (AoA) is unquestionable, selecting an appropriate beamwidth is also essential in the mmWave system networks. 
%In contrast to common belief, narrow beamwidth,...

Several measurements are already conducted in mmWave communications and prove that beamwidth has a critical effect on the channel parameters. In \cite{Manabe, Williamson,Kim}, antenna directivity (indirectly beamwidth) dependency to the delay and angle spread of the link is investigated at 28, 38 and 60 GHz. %These experiments show that delay and angle spread are proportional to beamwidth both for LOS and NLOS cases. 
\cite{Rappaport_aoa, Lee} conduct some outdoor experiments at 28 and 38 GHz with different beamwidth antennas and measure the incurred path loss. \cite{Rajagopal} provides similar outdoor LOS and NLOS tests and collect data of captured energy (received power) for several beamwidth values at 28 and 40 GHz. Both experiments show that wider beams have better performance, i.e. capture more energy and experience less path loss. In \cite{Dogan}, optimum beamwidth is measured in case of blockage occurs within the channel where wider beamwidths are provided based on beam expansion. %This is simply due to reason that beamforming of the channel, spatially, filters some information out too. 
While the nonnegligible effect of the beamwidth on mmWave communications is demonstrated with several other measurement results, on the other hand, very few beamwidth analyses on the performance metrics are proposed so far. In \cite{Va}, it is shown that there is an optimal non-zero beamwidth (around $5^{\circ}$) that maximizes the coherence time of the time-varying vehicular channel at 60 GHz. In \cite{Akoum}, analysis results show that $10^{\circ}$ beamwidth has better coverage, less interference compared to $30^{\circ}$ for mmWave cellular networks. A more related work \cite{Vakilian} studies AoA estimation error effects on bit-error-rate (BER) with different beamwidths for the clustered channel model. %However, the focus is on the AoA estimation errors and the optimum beamwidth problem is still not addressed. Furthermore, a reconfigurable antenna is taken into account in the analysis which is impractical in mmWave systems. 
Finally, a detailed analysis of the link between the channel angular dispersion and the antenna structure is given in \cite{Yang}. However, the channel is simply assumed to be Rician and no clustering approach is adopted as generally seen in mmWave channels. To the best of authors' knowledge, a detailed theoretical analysis of the beamwidth and received power relation in the case of misalignment for clustered mmWave channels is not studied. 

In this paper, we provide an analytical framework for the optimum beamwidth that maximizes the received power for indoor mmWave clusters, in the case of misalignment. We first give the relation between beamwidth and the captured power from the cluster. To do so, we use two different intra-cluster channel models, IEEE 802.11ad \cite{11ad} and our previous work, RT-ICM \cite{Yaman2}. Then, we combine it with the antenna gain at an arbitrary cluster AoA and provide an overall received power and beamwidth relation. In the analysis, we consider the uniform linear array (ULA) antenna type for the antenna gain, and two beam models to approximate the main lobe array pattern; rectangular and triangular. We show that when the misalignment error is smaller than the standard deviation of the cluster power spectrum, the optimum beamwidth is at zero, while the theoretical maximum received power approaches to a constant. For this case, we give practical limits of the optimum beamwidth with the relation to the number of elements such that sacrificing from the maximum received power in the order of tenths can reduce the required number of antenna elements significantly. However, when the alignment error is larger than the standard deviation, we show that the optimum beamwidth increases rapidly to a level larger than the alignment error. Finally, we evaluate the performance of the analysis by comparing the analytical results with simulations for an indoor mmWave cluster. The work we propose in this paper will give insights to the optimum antenna array design in both MIMO and massive MIMO applications for future mmWave systems network.

The rest of the paper is organized as follows. In Sec. \ref{system_model}, the overall problem is defined analytically and RT-ICM is summarized. In Sec. \ref{Antenna}, the relation between the beamwidth and the antenna structure is given. Sec. \ref{Angulardistribution} studies the beamwidth effect on the channel side. Sec. \ref{problemformulation} gives the problem formulation for two cluster models and provides simplified expressions for the optimum beamwidth. Performance evaluations of the extracted expressions are analyzed in Sec. \ref{performanceevaluation}. Finally, Sec. \ref{conclusion} concludes the paper.

\section{System and Channel Model} \label{system_model}

The optimum beamwidth problem has two sides; while decreasing beamwidth and increasing directivity of an array beam, (1) antenna gain increases, (2) captured energy from the cluster channel decreases. Hence, received power on the antenna terminals directly depends on the beamwidth. %In this paper, we investigate the theoretical limits of the maximum received power from a cluster while beamwidth approaches to zero. 

\begin{figure}[t]
\centering
\includegraphics[scale=0.52]{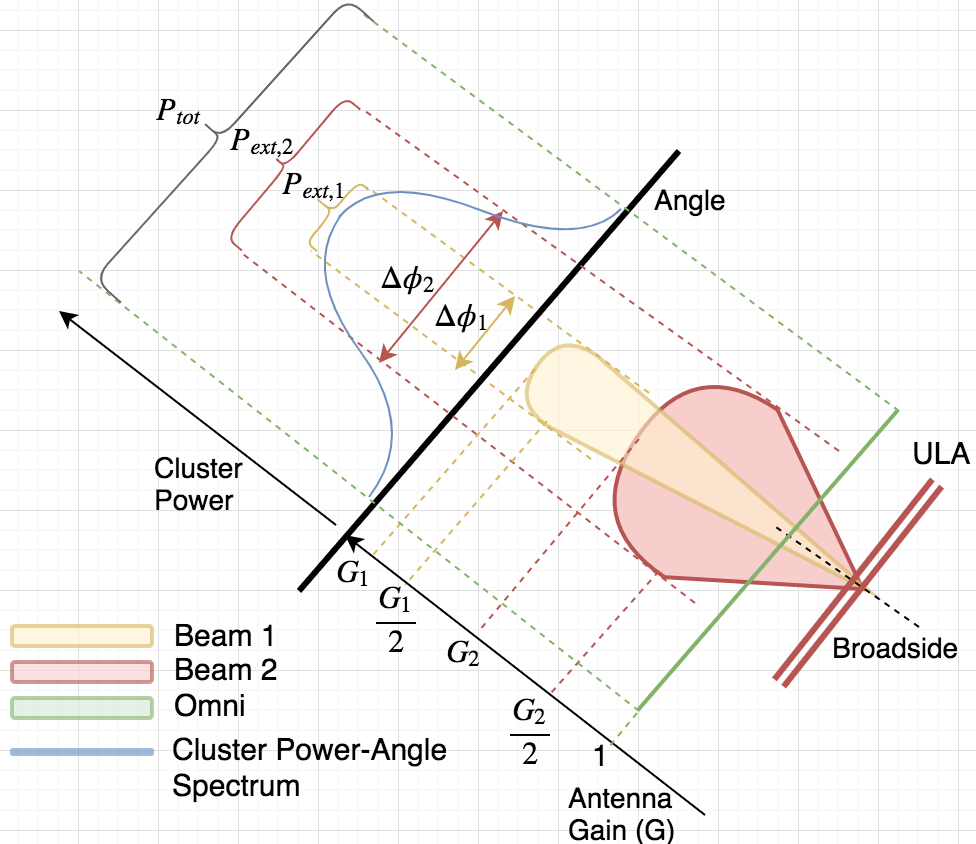} \\
\caption{Visualization of the optimum beamwidth problem at the receiver}\label{beamwidth}
\end{figure}
Received power on the antenna terminals is given in \cite{Orfanidis} as
\begin{equation}
P_R= \mathcal{P}_{inc}\frac{G\lambda^2}{4\pi} \label{PR}
\end{equation} 
where $\mathcal{P}_{inc}$ is the power density per \textit{area} front of the antenna; $G$ is the receiver antenna gain at maximum direction and $\lambda$ is the wavelength. Note that when an omnidirectional antenna is employed at the receiver, whose gain is $1$ at all directions, received power equals the available total cluster power. Then, the available power in front of the antenna can be given as

\begin{equation}
P_{tot}=\mathcal{P}_{inc}\frac{\lambda^2}{4\pi} \label{Ptot}
\end{equation}

Let $P_{ext}\leq P_{tot}$ is the extracted power from the cluster by a directional antenna. The equality holds whenever the beamwidth of the antenna covers entire cluster spatially\footnote{In this paper, we assume antenna beam models whose gain is 0 outside the beamwidth. That structure is discussed in Sec. \ref{Antenna}.}. Then, the received power given in Eq. (\ref{PR}) can be represented as a function of half-power beamwidth ($\Delta\phi$) and can be updated as
\begin{equation}
P_R(\Delta\phi)=G(\Delta\phi)P_{ext}(\Delta\phi) \label{PR_deltaphi}
\end{equation}

In Fig. \ref{beamwidth}, an example diagram of the discussion is illustrated with a comparison of two beams created by a ULA and steered towards a cluster AoA.  
%The effect of inverse relation between $\Delta\phi$ and $G_R$, and $P_{ext}$ can be seen from Eq. (\ref{PR_deltaphi}) immediately, as discussed in Sec. \ref{system_model}. In the next two sections, we give expressions for $G_R(\Delta\phi)$ and $P_{ext}(\Delta\phi)$.

Analysis of the problem requires the knowledge of spatial representation of the intra-cluster channel. However, while phased array antennas are well-studied in the literature and allow us to derive antenna gain-beamwidth relation, on the other hand, intra-cluster angular behavior of the mmWave channels is still not understood very well. In 3GPP channel model \cite{3GPP}, angular distribution of cluster power is simply modeled with a fixed number of rays with equal power levels. In 60 GHz WLAN standards IEEE 802.11ad \cite{11ad} and IEEE 802.11ay \cite{11ay}, a more intuitive model is adopted based on the measurements such that the power angular spectrum is distributed normally with $N(0, \sigma)$ where $\sigma=5$ for conference room and cubicle environments and $\sigma=10$ for living room channel models. Considering the site-specific nature of the mmWave channels, these models are likely to fail for different type of environments. %Hence, we introduced a simple ray tracing intra-cluster model for mmWave channels in our previous work \cite{Yaman2} that gives accurate results for any stationary environments. %In the next subsections, we summarize the model first and then express the research statement analytically by exploiting the adopted channel model.
	
	\subsection{Ray Tracing based Intra-Cluster Channel Model \cite{Yaman2}} \label{ICM}

\begin{figure}[t]
\centering
\includegraphics[scale=0.5]{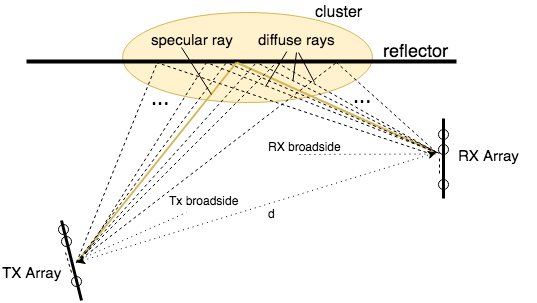} \\
\caption{First-order reflection cluster model of the RT-ICM}\label{clusterdef}
\end{figure}

In \cite{Yaman2}, we introduce a mmWave intra-cluster model based on ray-tracing (RT-ICM) that takes only first-order reflections into account. The general first-order reflection cluster definition of the model is illustrated in Fig. \ref{clusterdef}. In our model, we also add the scattering effect based on the material properties. It outputs the power distribution both in angle and time domain within the cluster and can be used for both indoor and outdoor mmWave systems in any type of stationary environments. In specific, the proposed cluster channel model comprises three main steps; namely, basic geometry modeling, theoretical cluster modeling, and binned channel impulse response generation. First, Basic Geometric Model (BGM) takes first-order reflection cluster environment parameters as inputs and generates the supported angle spread (SAS), $S_{\phi}=\phi_{max}-\phi_{min}$, at the receiver. Second, running the BGM for each infinitely large number of rays that are within the supported angle spread; angle, time, power and phase of the rays are calculated; thereby generating the theoretical cluster impulse response. The output theoretical baseband cluster impulse response (TC-CIR) is given by \cite{Yaman2}
\begin{equation}
\begin{split}
c_T(t_{sp},\phi_{sp})&=a_{sp} e^{j\varphi_{sp}}\delta (t_{sp})\delta (\phi_{sp})\\
&+\sum_{k=0}^{N_r^d-1} a_k e^{j\varphi_k} \delta (t_{sp}-\tau_k) \delta (\phi_{sp}-\alpha_k) \label{CIR_theo}
\end{split}
\end{equation}
where $a_{sp}$, $\varphi_{sp}$, $t_{sp}$ and $\phi_{sp}$ are the amplitude, phase, time of arrival (ToA) and AoA of the specular ray, respectively; $a_{k}$, $\varphi_{k}$, $\tau_{k}$, $\alpha_{k}$ are amplitude, phase, delay, offset AoA of the $k$-th ray, respectively. $\delta (.)$ is  Dirac delta function and $N_r^d$ is the number of rays.

\begin{figure}[t]
\centering
\includegraphics[scale=0.45]{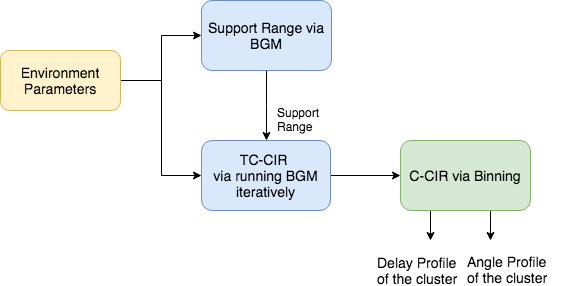} \\
\caption{Flowchart diagram of the cluster channel impulse response generation.}\label{flowchart}
\end{figure}

\begin{figure}[t]
\centering
\includegraphics[scale=0.23]{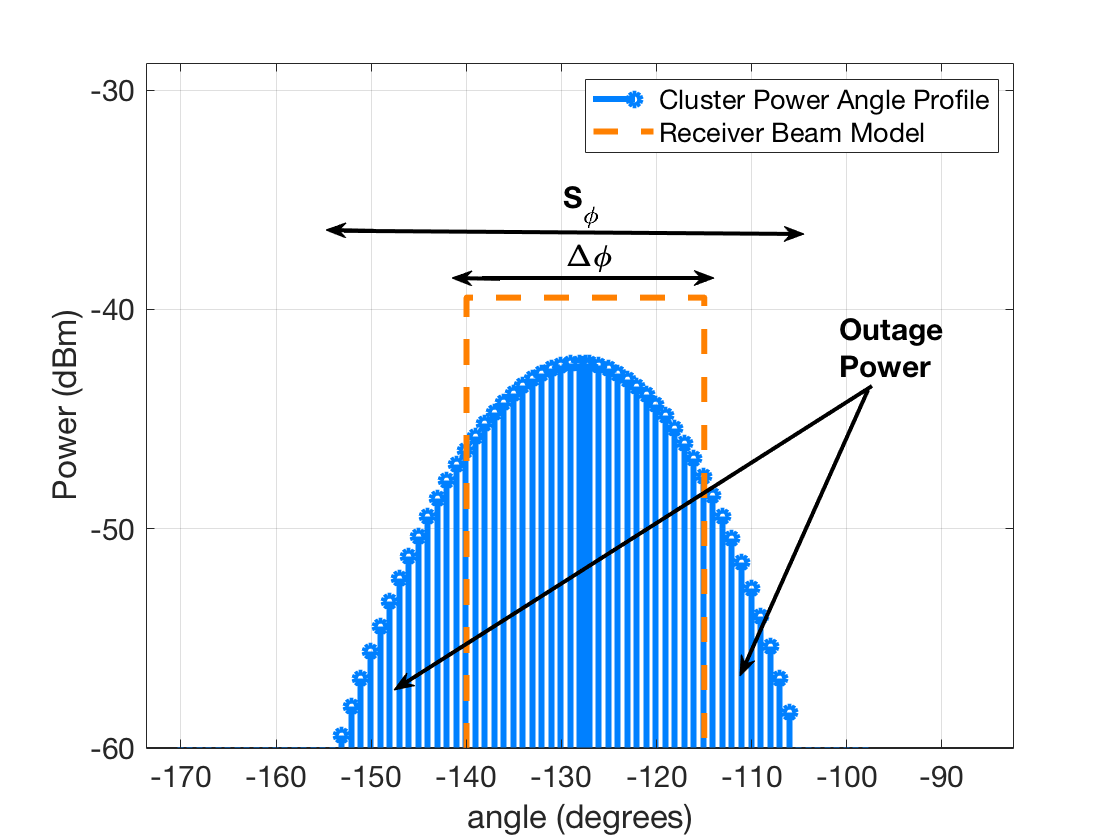} \\
\caption{An example cluster power angle profile of RT-ICM.}\label{outage_figure}
\end{figure}

Finally, the theoretical impulse response is binned in the angle and time domain according to the bandwidth and angle resolution of the communication system. Basically, the binned version is the impulse response that the receiver \textit{sees}. We simply call it cluster channel impulse response (C-CIR). The diagram of the overall modeling process is given in Fig. \ref{flowchart}. Note that, since we are interested in estimating the received power at the antenna terminals in this paper, i.e. before receiver signal processing, we will use the theoretical impulse response given in Eq. (\ref{CIR_theo}) as the cluster power angle profile. Then, the total power in the cluster is given as
\begin{equation}
P_{tot}=a_{sp}^2 + \frac{S_{\phi}}{N_r^d} \sum_{k=0}^{N_r^d} a_k^2 \label{totalpower}
\end{equation}
where $S_{\phi}/N_r^d$ term is inserted for the integral approximation. An example of a cluster angle profile output of RT-ICM is displayed in Fig. \ref{outage_figure} for $N_r^d=75$ and $S_{\phi}=75^{\circ}$.

\section{Antenna Structure and Gain} \label{Antenna}

\begin{figure}[t]
\centering
\includegraphics[scale=0.55]{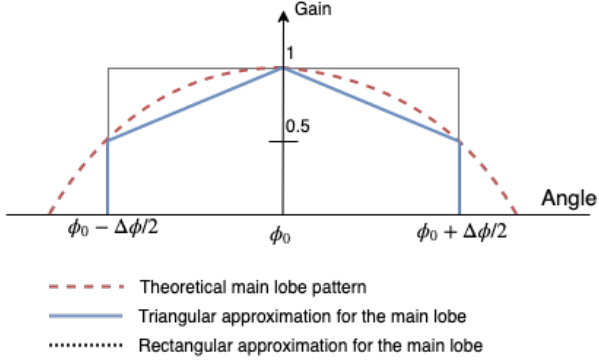} \\
\caption{Antenna pattern models considered in the paper.}\label{patternapprox}
\end{figure}

As Eq. (\ref{PR_deltaphi}) suggests, antenna gain is the counterpart of the captured cluster power in the equation for a certain beamwidth. Finding an expression for antenna gain as a function beamwidth for an arbitrary array design is not easy and out-of-scope of the paper. Instead, we adopt the well-known uniform linear array (ULA) design where the spacings between the elements are equal and we seek to find the relationships between antenna gain, beamwidth, number of elements and scan angle for a ULA. 

\subsection{Beam Pattern Model} \label{beampatternmodel}
In this paper, we use two models for the beam pattern; a rectangular window and a triangular window, both are seen in Fig. \ref{patternapprox} for a steering (scan) angle of $\phi_0$. Both approximations ignore the sidelobes; thereby modeling only the main lobe. Expressions of the shown functions for rectangular and triangular model, respectively, are,
\begin{align}
W_{R}= &
\begin{cases} \label{WR}
1,  & \phi_0-\Delta\phi/2 > \phi > \phi_0+\Delta\phi/2 \\
0, & otherwise
\end{cases}\\ 
W_{T}= &
\begin{cases} \label{WT}
1-\frac{|\phi-\phi_0|}{\Delta\phi},  & \phi_0-\Delta\phi/2 > \phi > \phi_0+\Delta\phi/2 \\
0, & otherwise
\end{cases}
\end{align}

It can be seen that the rectangular model amplifies the signal at the beam edges while the triangular model 
attenuates. From that perspective, one can consider the rectangular model as the upper bound whereas the triangular model as the lower bound.
 
\subsection{Antenna Gain vs. Beamwidth}
Since the efficiency of phased array antennas are almost perfect \cite{Hansen}, we use antenna gain and directivity interchangeably, i.e. $G=D$. Then, for uniform excitation of the elements\footnote{Or rectangular window tapering for window-based array designs}, antenna gain at an arbitrary scan angle $0^{\circ}<\phi_0<180^{\circ}$ for ULA, with an inter-element spacing of $d=\lambda/2$, is given as \cite{Hansen}
%\begin{equation*}
%G=\frac{N^2}{N+2\sum_{n=1}^{N-1}(N-n) \text{sinc}(n\frac{2\pi}{\lambda}d) \cos (n\frac{2\pi}{\lambda}d \sin \phi_0)}
%\end{equation*}
%where $N$ number of antenna elements; $d$ is the inter-element spacing in $\lambda$. Note that, for a special case of $d=\lambda/2$, equation reduces to following, independent of $\phi_0$:
\begin{equation}
G=N \label{GeqN}
\end{equation}
where $N$ is the number of antenna elements. Eq. (\ref{GeqN}) simply states that the antenna gain equals  the number of elements and is independent of scan angle for a fixed number of array elements. However, the beamwidth changes with scan angle and is given in degrees for $d=\lambda/2$ as \cite{Orfanidis, Balanis}
\begin{equation}
\Delta\phi=\frac{101.5^{\circ}}{N\sin \phi_0} \label{beamwidth_eq}
\end{equation}
for $0^{\circ}<\phi_0<180^{\circ}$. At exactly endfire, i.e. $\phi_0=0^{\circ},180^{\circ}$, $\Delta\phi=152.53^{\circ}/\sqrt{N}$.

%\begin{figure}[t]
%\centering
%\includegraphics[scale=0.22]{gainvsbmw.png} \\
%\caption{Antenna gain and beamwidth relation for different scan angles.}\label{gainvsbmw}
%\end{figure}

Combining Eq. (\ref{GeqN}) and (\ref{beamwidth_eq}), the ULA antenna gain  as a function of beamwidth, with uniform weighting, for $d=\lambda/2$, can be given as
\begin{equation}
G=\frac{101.5}{\Delta\phi \sin \phi_0} \label{Gvsdelta}
\end{equation}
for $0^{\circ}<\phi_0<180^{\circ}$ and the endfire gain is given as $G=(152.53^{\circ}/\Delta\phi)^2$. %Fig. \ref{gainvsbmw} plots Eq. (\ref{Gvsdelta}) for broadside, endfire and at $\phi=145^{\circ}$ radiation. %To get the insight, consider the following example: antenna gain variation for a fixed $\Delta\phi=20^{\circ}$ is $10$ dB over the entire azimuth scan angle. That shows how sensitive the gain to scan angles if $\Delta\phi_0$ is forced to be fixed.  
%As a result, since we define the antenna gain as the amount of amplification of the captured power, we set $G_R(\Delta\phi)=G$.

\section{Angular Distribution of Total Power and Extracted Power} \label{Angulardistribution}
In this section, we give expressions for the extracted power by a $\Delta\phi$-beamwidth antenna ($P_{ext}(\Delta\phi)$) using the power-angle spectrum of two cluster channel models, IEEE 802.11ad  and RT-ICM. To be comparable, we use the same total power, $P_{tot}$, obtained from RT-ICM in Eq. (\ref{totalpower}) for both cluster models. 
	\subsection{Extracted Power for IEEE 802.11ad Cluster Model}
Letting $\phi_{cl}$ is the cluster AoA, p.d.f. of the angular distribution of a cluster is given in \cite{11ad} as $f(\phi)=1/(\sqrt{2\pi\sigma^2)}e^{-(\phi-\phi_{cl})^2/(2\sigma^2)}$. Normalized power that is captured by the beam that is steered to $\phi_0$ with a beamwidth $\Delta\phi$  can be obtained by 
\begin{equation}
\rho=\int_{\phi_0-\Delta\phi/2}^{\phi_0+\Delta\phi/2} W(\phi) f(\phi) \text{d}\phi \numberthis \label{p}
\end{equation}
%\frac{\Delta\phi}{2}}

where $W(\phi)$ is the beam shape defined in Sec. \ref{beampatternmodel},  $W(\phi)=W_R$ for rectangular and $W(\phi)=W_T$ for triangular model. 

Then the \textit{average} extracted power from the cluster for IEEE standard model becomes
\begin{equation}
P_{ext}^{st}= P_{tot} \rho  \label{Pextst}
\end{equation}

%In the simulation results section, we set the number of rays to $N_r^d$, i.e. a large number, although the model given in \cite{11ad} is for only $14$ rays. That change doesn't affect the average results. On the other hand,  $14$ rays are observed after processing of received signal with a limited spatial resolution in the standard. However, in here, we intend to estimate the power before any processing occurs. Also, linear processing of the signal won't change the spatial pattern except that the observed angle spread might be lower than the channel real angle spread due to the filtering out the rays that are under the receiver sensitivity for processing. That effect will be described in Sec. \ref{performanceevaluation} in details.

	\subsection{Extracted Power for RT-ICM}
In RT-ICM, a cluster is generated totally by a single-order reflection and a reflector creates only one cluster. With this assumption, the strongest component in a cluster is the specular reflection which yields that the specular ray AoA can be set as the cluster AoA, i.e. $\phi_{sp}=\phi_{cl}$.  

In \cite{Yaman2}, rays within the angle spread are equally-separated with a fixed spacing, $\Delta\alpha$. Then, the number of rays that drops within $\Delta\phi$ is $N_b=\lfloor\Delta\phi/\Delta\alpha \rfloor$. Then, the extracted power for RT-ICM can be given in the same discrete angle domain as
\begin{equation}
P_{ext}^{rt}=a_{sp}^2 + \frac{\Delta\phi}{N_b} \sum_{k=m}^{m+N_b-1} a_k^2 \label{PextICM}
\end{equation}

where $m=arg\{a_k | a_k=\phi_0-\Delta\phi/2 + \text{d}\phi, k=0, 1, \hdots, N_r^d\}$ and $\text{d}\phi \geq 0$ is the minimum continuous angle that requires to select the first ray within the beamwidth. 

The optimization of the beamwidth requires to take the derivation of Eq. (\ref{PextICM}) with respect to $\Delta\phi$. However, $\Delta\phi$ is in the argument which makes the derivation hard. Instead, we give an alternative approach via approximation.

\paragraph{Gaussian Approximation}
In \cite{Yaman2}, the diffuse scattering loss is modeled with a raised cosine function which results in a \textit{Gaussian-like} received power spectrum in the spatial domain in the linear scale. Hence, we seek to fit a Gaussian function to the power angle profile output of RT-ICM. The Gaussian function is of the form $g(\phi)=ue^{(\phi-x)^2/v^2}$ where the parameters $u$, $v$ and $x$ control the peak, the width and the mean of the function, respectively. We select these parameters such that the resultant function fits best to the data. An algorithm is proposed for Gaussian fitting in \cite{Guo}. Then, the extracted power with a beamwidth $\Delta\phi$ is obtained by 

\begin{equation}
P_{ext}^{rt}=\int_{\phi_0-\Delta\phi/2}^{\phi_0+\Delta\phi/2} W(\phi)  g(\phi) \text{d}\phi \label{PextICM_gauss}
\end{equation}

%Note that we do not model $\phi$ as a random variable here, so Eq. (\ref{gphi}) is not a p.d.f. It is a fitting function of the data. In that sense, the output of the integral in Eq. (\ref{PextICM_gauss}) is in power units and differs from the integral in Eq. (\ref{p}) whose output is a probability. 

% 	\subsection{Outage Rate} \label{outage}
%Now that the total cluster power and the extracted power by a certain beamwidth is known, a metric to measure the outage at the cluster can be introduced. We define the \textit{outage power} for the cluster power that remains outside the 3dB beamwidth of the antenna as in Fig. \ref{outage_figure}. Then, the \textit{outage rate} that defines the rate of outage over total power is 
%\begin{equation}
%\gamma=\frac{P_{tot}-P_{ext}}{P_{tot}}=1-\frac{P_{ext}}{P_{tot}}
%\end{equation}

%Then, from Eq. (\ref{Pextst}) and (\ref{PextICM_gauss}), 802.11ad model and RT-ICM outage rates are defined as follows, respectively:
%\begin{equation}
%\gamma^{st}=1-\rho
%\end{equation}
%\begin{equation}
%\gamma^{icm}=1-\frac{P_{ext}^{icm}}{P_{tot}}
%\end{equation}

\section{Problem Formulation} \label{problemformulation}
%Now that we define antenna gain and channel power in terms of the beamwidth, next we find the optimum beamwidth that maximizes the received power in Eq. (\ref{PR_deltaphi}). 
In case of an imperfect channel knowledge, cluster angle of arrival, $\phi_{cl}$, is estimated with an error. Additionally, a quantization error is introduced at the receiver when a codebook is implemented for beamforming. Then the total misalignment error at the receiver can be defined as\footnote{Absolute value is inserted to keep the error positive as all the functions used in the analysis are symmetric.} $\delta=|\phi_0-\phi_{cl}|$. We first set up the problem formulation with $W(\phi)=W_R$.

%Note that from Eq. (\ref{Gvsdelta}), the antenna gain goes to infinity while beamwidth approaches zero. On the other hand, extracted cluster power given in Eq. (\ref{Pextst}) and (\ref{PextICM_gauss}) goes to $-\infty$ with reducing beamwith. This suggests that the Eq. (\ref{PR_deltaphi}) converges to a limit while $\Delta\phi \rightarrow 0$. Apparently, zero beamwidth is impractical and extremely small beamwidth comes with an expense of inapplicably increased number of antenna elements. In the following two subsections, we give the problem setup for two aforementioned cluster models and prove that maximum received power is at $\Delta\phi = 0$. Then, we provide an expression for the theoretical maximum received power by performing the asymptotic analysis of the received power equation and give analytical beamwidth solutions to the practical limits.

	\subsection{IEEE 802.11ad} \label{problem_st}

Plugging antenna gain and extracted power equations in Eq. (\ref{Gvsdelta}) and (\ref{Pextst}), respectively, into Eq. (\ref{PR_deltaphi}),

\begin{equation}
P_R(\Delta\phi)=\frac{101.5^{\circ}}{\Delta\phi \sin \phi_0}P_{tot}\rho \label{PR_st}
\end{equation}
where $\rho$ is integral given in Eq. (\ref{p}). From \cite{Papoulis}, the integration of a Gaussian can be defined with the error function (\textit{erf}). Then,
\setlength{\jot}{10pt} %for the vertical space between lines
\begin{align*}
\rho&= \int_{\phi_{cl}+\delta-\Delta\phi/2}^{\phi_{cl}+\delta+\Delta\phi/2} W_R(\phi) \frac{1}{\sqrt{2\pi\sigma^2}}e^{-\frac{(\phi-\phi_{cl})^2}{2\sigma^2}} \text{d}\phi \\
&=\frac{1}{2} \left( \text{erf} \left( \frac{\Delta\phi + 2\delta}{2\sqrt{2}\sigma} \right) + \text{erf} \left( \frac{\Delta\phi - 2\delta}{2\sqrt{2}\sigma} \right) \right) \numberthis \label{rho_imperfect}
\end{align*}
where $erf(z)=1\sqrt{2\pi} \int_0^z e^{-y^2/2}dy$. Plugging into Eq. (\ref{PR_st}), the received power can be given as following:
\begin{align*}
P_R&(\Delta\phi)=\\
&\frac{50.75P_{tot}}{\Delta\phi \sin \phi_0} \left( \text{erf} \left( \frac{\Delta\phi + 2\delta}{2\sqrt{2}\sigma} \right) + \text{erf} \left( \frac{\Delta\phi - 2\delta}{2\sqrt{2}\sigma} \right) \right) \numberthis \label{PR_st_b4_deriv}
\end{align*}
such that $0 < \Delta\phi$. %Note that $erf(z)$ is defined in the $(0,1]$ range for $z>0$ and starts saturating 1 for $z\geq 2$. Since $erf(z)$ is linear in the range of approximately $(0,0.5]$, multiplying it with $1/z$ mirrors the function on the z axis which now saturates for smaller values of $z$. That proves the statement that the maximum received power is obtained while $\Delta\phi \rightarrow 0$.  

\subsubsection{Maximization of $P_R$} \label{PR_maximization_imperfect}

We seek to find the optimum $\Delta\phi=\Delta\phi_{opt}$ that maximizes the Eq. (\ref{PR_st_b4_deriv}). Since $P_{tot}$ and $\sin \phi_0$ are positive and scaling doesn't affect the optimization, the problem can be reduced to the following:

\begin{equation}
max \ \frac{\text{erf} \left( \frac{\Delta\phi + 2\delta}{2\sqrt{2}\sigma} \right) + \text{erf} \left( \frac{\Delta\phi - 2\delta}{2\sqrt{2}\sigma} \right)}{\Delta\phi} \label{maxPR}
\end{equation}

Making an argument whether the Eq. (\ref{maxPR}) is concave is not straightforward. To investigate the concavity of the function, we apply Second Derivative Test which is summarized as follows. If $q'(z_0)=0$ and $q''(z_0)<0$, then $z_0$ is the local maximum of the function $q(z)$. The simplified conditions derived from the first and the second derivatives of the Eq. (\ref{maxPR}) are given in Eq. (\ref{eqn_dbl_x}) and (\ref{eqn_dbl_y}) at the top of page 6, respectively. The details are placed in Appendix \ref{maxPRmisalign}.

\begin{figure*}[!t]
% ensure that we have normalsize text
\normalsize
% Store the current equation number.
\setcounter{mytempeqncnt}{\value{equation}}
% Set the equation number to one less than the one
% desired for the first equation here.
% The value here will have to changed if equations
% are added or removed prior to the place these
% equations are referenced in the main text.
%\setcounter{mytempeqncnt}{21}
\begin{equation}
\label{eqn_dbl_x}
\Delta\phi \left( e^{-(\Delta\phi+2\delta)^2/8\sigma^2} + e^{-(\Delta\phi-2\delta)^2/8\sigma^2} \right) -\sigma \sqrt{2\pi} \left( \text{erf} \left( \frac{\Delta\phi + 2\delta}{2\sqrt{2}\sigma} \right) + \text{erf} \left( \frac{\Delta\phi - 2\delta}{2\sqrt{2}\sigma} \right)  \right)=0
\end{equation}
\begin{equation}
\label{eqn_dbl_y}
\frac{\Delta\phi(\Delta\phi(\Delta\phi+2\delta)+8\sigma^2)}{e^{(\Delta\phi+2\delta)^2/8\sigma^2} }+\frac{\Delta\phi(\Delta\phi(\Delta\phi-2\delta)+8\sigma^2)}{e^{(\Delta\phi-2\delta)^2/8\sigma^2} } > \left( \text{erf} \left( \frac{\Delta\phi + 2\delta}{2\sqrt{2}\sigma} \right) + \text{erf} \left( \frac{\Delta\phi - 2\delta}{2\sqrt{2}\sigma} \right) \right)
\end{equation}
% Restore the current equation number.
%\setcounter{equation}{\value{mytempeqncnt}}
% IEEE uses as a separator
\hrulefill
% The spacer can be tweaked to stop underfull vboxes.
\vspace*{4pt}
\end{figure*}

 \begin{figure}[t]
\centering
\includegraphics[scale=0.23]{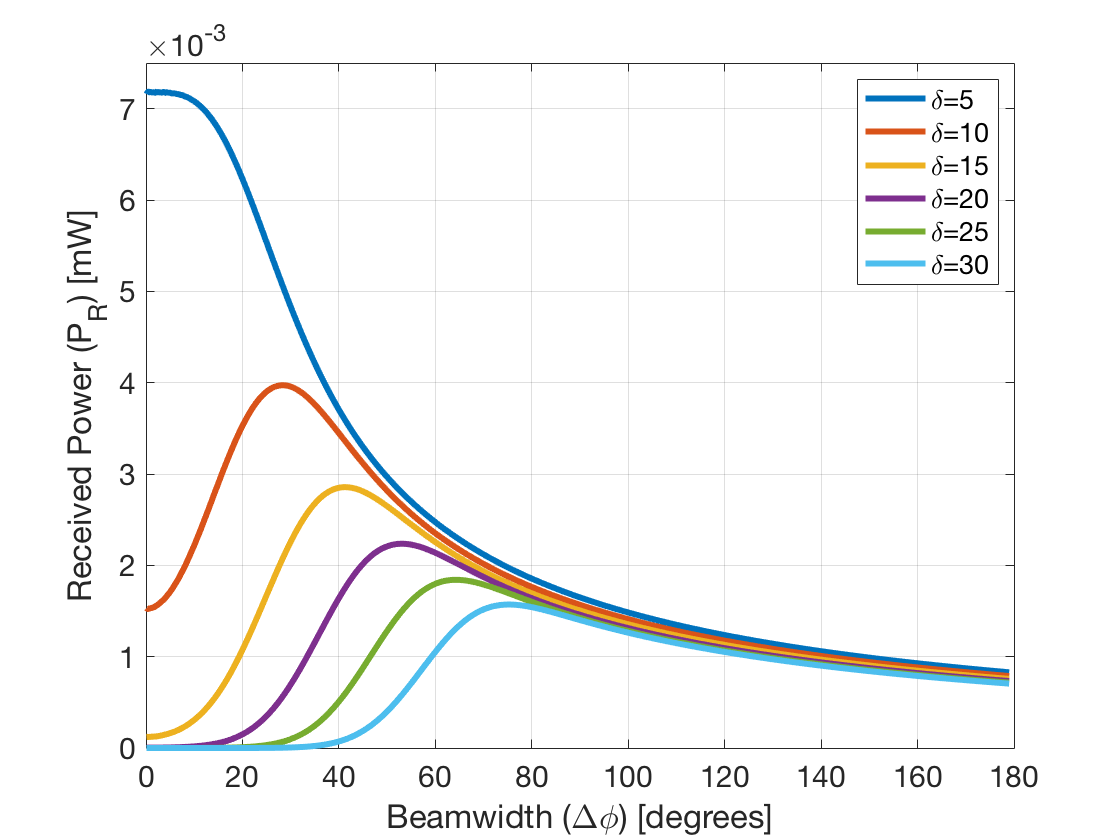} \\
\caption{Received power in case of misalignment for $\sigma=5$ and $P_{tot}=1.2\mu W$.}\label{PRimperfect}
\end{figure}

Due to the \textit{erf} function, no closed-form is available for Eq. (\ref{eqn_dbl_x}), thus, solving it for $\Delta\phi>0$ is performed numerically, for fixed $\delta$ and $\sigma$. The function has always 2 roots, one at the infinity. Plugging the roots to Eq. (\ref{eqn_dbl_y}), it can be seen that the condition fails for the infinity root. Furthermore, it is counter-intuitive to have the infinitely large $\Delta\phi_{opt}$ for the maximized received power. In fact, this root is the local minimum of the function which \textit{minimizes} the received power. 

An example simulation of the $P_R$ in linear scale with respect to $\Delta\phi$ for different $\delta$ values are given in Fig. \ref{PRimperfect} where $P_{tot}=1.2 \mu W$, $\sigma=5^{\circ}$ and $\phi_0=53^{\circ}$.
It can be seen that the second derivative test would result in the single local maximum, i.e., global maximum. This leads to the following remark.

\paragraph*{Remark 1} \label{Remark1}
There exists a unique $\Delta\phi>0$ that satistifies the both conditions in Eq. (\ref{eqn_dbl_x}) and (\ref{eqn_dbl_y}) for the given $\delta$ and $\sigma$ such that $\delta > \sigma$. This unique value \textit{is} the optimum beamwidth, $\Delta\phi_{opt}$, that maximizes the received power for the given $\delta$ and $\sigma$. 

Also, to be proven in the simulation results in Sec. \ref{performanceevaluation}, we propose another important remark:

\paragraph*{Remark 2} \label{Remark2}
The following statements are always true. (1) When $\delta > \sqrt{2}\sigma$, $\Delta\phi_{opt}>2\delta$, (2) When $\delta \leq \sigma$, $\Delta\phi_{opt}=0$.
\\

\begin{figure}[t]
\centering
\includegraphics[scale=0.45]{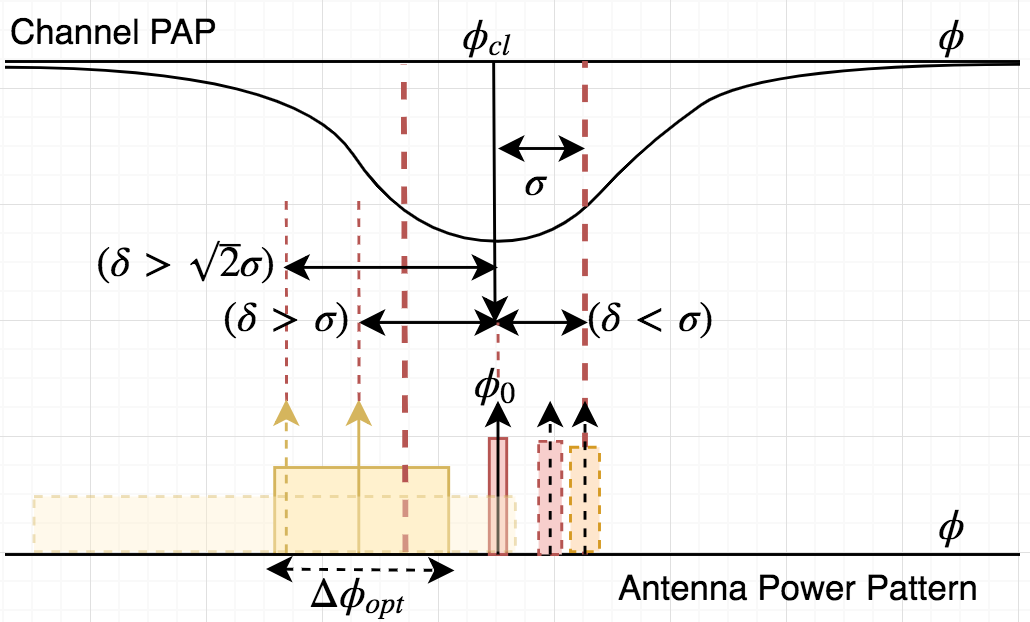} \\
\caption{Optimum beamwidth in case of misalignment.}\label{misalignment}
\end{figure}

To get the insight of the results, consider the diagram given in Fig. \ref{misalignment}. The relation between $\delta$ and $\sigma$ has a significant effect on the value of $\Delta\phi_{opt}$. As long as $\delta \leq \sigma$, it is enough for the antenna to capture a little from the first $\sigma$ region of the channel spectrum and antenna gain dominates the maximum received power. However, once $\delta>\sigma$, channel gain that is captured from the tail is considerably small such that even the antenna gain cannot tolerate. Hence, regardless of $\delta$, the maximum received power is achieved only when the beam captures from the first $\sigma$ region of the channel spectrum. Furthermore, if $\delta>\sqrt{2}\sigma$, beam should cover the center of the cluster. That is, $\Delta\phi_{opt}/2 > \delta$, or $\Delta\phi_{opt} > 2\delta$.

Finally, once optimum beamwidth is determined, maximum received power can be obtained by plugging $\Delta\phi_{opt}$ into Eq. (\ref{PR_st_b4_deriv}),

\begin{equation}
P_{max}=P_R(\Delta\phi_{opt}) \label{Pmaximperfect}
\end{equation}

\subsection{RT-ICM} \label{RTICM_imperfect}

Apparently, the equations, given in Eq. (\ref{Pextst}) and (\ref{PextICM_gauss}), are equal with different notations. Hence, skipping the intermediate steps, received power for RT-ICM is given as
\begin{align*}
P_R^{rt}&(\Delta\phi)=\\
&\frac{50.75u v \sqrt{\pi}}{\Delta\phi \sin \phi_0} \left( \text{erf} \left( \frac{\Delta\phi + 2\delta}{2v} \right) + \text{erf} \left( \frac{\Delta\phi - 2\delta}{2v} \right) \right) \numberthis \label{PR_rt_b4_deriv}
\end{align*}

Maximization procedure of $P_R^{rt}(\Delta\phi)$ is analogous to the 802.11ad case. In fact, setting $\sigma=v/\sqrt{2}$ in Eq. (\ref{eqn_dbl_x}) and (\ref{eqn_dbl_y}), conditions for RT-ICM can be provided. 

Note that the IEEE 802.11ad standard sets $\sigma$ to a fixed value for specific environments whereas RT-ICM generates different $v$ for each cluster. In any way, after getting $v$, one can convert it to $\sigma$ and use the same notation with 802.11ad. 

\subsection{Asymptotic Analysis of Perfect Alignment} \label{asymptotic_perfect}
It is already shown that $\Delta\phi_{opt}=0$ when $\delta \leq \sigma$. Note that infinitely many elements ($N=\infty$) are required to achieve $\Delta\phi_{opt}=0$, which is impractical. In this section, we perform an asymptotic analysis of the maximum achievable received power while $\Delta\phi_{opt}\to 0$ and provide an expression that results in practical beamwidth values that captures a percentile of the maximum achievable power. We perform the analysis with the 802.11ad model but give the counterpart expressions for RT-ICM at the end of the section as well.

To simplify the analysis and create a generic expression for any $\sigma$, let us assume $\delta=0$, i.e. $\phi_0=\phi_{cl}$. Then from Eq. (\ref{p}), for $W(\phi)=W_R$, $\rho$ becomes\footnote{It can also be verified by plugging $\delta=0$ in Eq. (\ref{rho_imperfect}).},

\begin{equation}
\rho=\text{erf} \left( \frac{\Delta\phi}{2\sqrt{2}\sigma} \right)
\end{equation}

Plugging into Eq. (\ref{PR_st}), received power equation simplifies to,

\begin{equation}
P_R(\Delta\phi)=\frac{101.5P_{tot}}{\Delta\phi \sin \phi_0} \text{erf} \left( \frac{\Delta\phi}{2\sqrt{2}\sigma} \right) \label{PR_st_perfect}
\end{equation}

		\subsubsection{Maximum Received Power}
We now seek to find the maximum received power as the limit $\Delta\phi \rightarrow 0$. Eq. (\ref{PR_st_perfect}) is in the $0/0$ indeterminate form for $\Delta\phi =0$. Applying L'Hopital rule, the maximum achievable received power is 
 \begin{equation}
 P_{max}^t=P_R(0)=\frac{40.5P_{tot}}{\sigma \sin \phi_0} \label{Pmax_st}
 \end{equation}
 Intermediate steps are given in the Appendix \ref{802.11ad}.
 
		\subsubsection{Optimum Practical Beamwidth} \label{11ad_opt_beam}
Note that the maximum received power given in the Eq. (\ref{Pmax_st}) is theoretical. Achieving very small beamwidth requires impractically high number of antenna elements as the relation given in Eq. (\ref{beamwidth_eq}) suggests. However, we can keep the beamwidth in practical ranges while sacrificing subtle from the received power. 

Let $0<\eta\leq 1$ be the coefficient such that 
\begin{equation}
P_{\eta}^{t}=\eta P_{max}^{t} \label{Peta}
\end{equation} 
where $P_{\eta}^{t}$ is the $\eta$-percentile power of the $P_{max}^{t}$. Then, from Eq. (\ref{PR_st_perfect}),
\begin{equation}
\frac{101.5P_{tot}}{\Delta\phi \sin \phi_0} \text{erf} \left( \frac{\Delta\phi}{2\sqrt{2}\sigma} \right)=P_{\eta}^{st}=\eta \frac{40.5P_{tot}}{\sigma \sin \phi_0}
\end{equation}

Simplifying the equation and setting $\Delta\phi=\Delta\phi_{\eta}$, i.e. practical $\eta$-percentile beamwidth,
\begin{equation}
\frac{\Delta\phi_{\eta}}{\text{erf} \left( \frac{\Delta\phi_{\eta}}{2\sqrt{2}\sigma} \right)}  =  \frac{2.5\sigma}{\eta} \label{deltaerf}
\end{equation}

Similar to the imperfect alignment case in Eq. (\ref{maxPR}), Eq. (\ref{deltaerf}) shows that the optimum beamwidth depends only on the angle spread of the cluster while total cluster power and the scan angle has no effect at all. However, different from the imperfect alignment, an approximate solution can be provided for  Eq. (\ref{deltaerf}). Expanding the $erf(z)$ function to Taylor series such that $\text{erf}(z)=2\left( z- z^3/3 + z^5/10 - z^7/42 + \hdots \right)/\sqrt{\pi}$ where $z=\Delta\phi/\sqrt{8}\sigma$, we can approximate it ignoring the high order terms when $\Delta\phi/\sqrt{8}\sigma<1$. Referring to Appendix \ref{802.11ad} for the derivation, the $\eta$-percentile beamwidth can be given, approximately,
\begin{equation}
\Delta\phi_{\eta} \approx 4.89 \ \sigma \sqrt{1-\eta} \label{beamwidth_11ad}
\end{equation}

As shown in the simulation results section, Eq. (\ref{beamwidth_11ad}) provides a strong yet simple tool to determine the optimum beamwidth within the practical limits. However, once $\Delta\phi > \sqrt{8}\sigma$, approximation of the error function starts getting apart from the real value. Setting $\Delta\phi \leq \sqrt{8}\sigma$ as the support region of $\Delta\phi$, from Eq. (\ref{beamwidth_11ad}),
\begin{align*}
4.89 \ \sigma \sqrt{1-\eta} &\leq \sqrt{8}\sigma \\
\eta & \geq 0.667 \numberthis \label{eta}
\end{align*}
which provides the lower bound accuracy limit of the Eq. (\ref{beamwidth_11ad}). Finally, to prove that Eq. (\ref{beamwidth_11ad}) results in practical beamwidth values for conference room environments, from Eq. (\ref{eta}) and (\ref{beamwidth_11ay_conf}), beamwidth is given in the support range of $\Delta\phi_{\eta} \leq 14.10$ which can be achieved at the broadside using $N\geq 8$ antenna elements.

Plugging the given standard deviation for conference room environments \cite{11ad} $\sigma=5$, Eq. (\ref{Pmax_st}), (\ref{deltaerf}) and (\ref{beamwidth_11ad}) becomes 
 \begin{equation}
 P_{max}^{st} =\frac{8.1P_{tot}}{\sin \phi_0} \label{Pmax_conf}
 \end{equation}
\begin{equation}
\frac{\Delta\phi^{st}}{\text{erf} \left( \frac{\Delta\phi^{st}}{10\sqrt{2}} \right)}  =  \frac{12.5}{\eta} \label{deltaerf_conference}
\end{equation}
\begin{equation}
\Delta\phi^{st}= 24.45 \sqrt{1-\eta} \label{beamwidth_11ay_conf}
\end{equation}

Due to the similar discussion made in Sec. \ref{RTICM_imperfect}, we just give the RT-ICM counterpart expressions here. With notation changes, received power equation becomes
\begin{equation}
P_R^{rt}(\Delta\phi)=\frac{101.5^{\circ} uv\sqrt{\pi}}{\Delta\phi \sin \phi_0} \text{erf} \left( \frac{\Delta\phi}{2v} \right) \label{PRdelta_icm}
\end{equation}

The maximum received power for the RT-ICM is
 \begin{equation}
 P_{max}^{rt} =\frac{101.5u}{\sin \phi_0} \label{Pmax_icm}
 \end{equation}		

Exact equation for the practical $\eta$-percentile beamwidth, $\Delta\phi_{\eta}$, for the RT-ICM,

\begin{equation}
\frac{\Delta\phi^{rt}}{\text{erf} \left( \frac{\Delta\phi^{rt}}{2v} \right)} =\frac{v\sqrt{\pi}}{\eta } \label{deltaphi_icm}
\end{equation}

and its approximation for $\Delta\phi \leq 2v$,
\begin{equation}
\Delta\phi^{rt} \approx 3.46 \  v \sqrt{1-\eta} \label{deltaphi_icm_approx}
\end{equation}

%whose derivations are given in Appendix \ref{ICM_appen}.

\begin{table}[!t]
\centering
\caption{Asymptotic Analysis of $P_{max}$ and Performance Comparison of 802.11ad and RT-ICM} 
\label{asym}
\begin{tabular}{|c||c|c||c|c||}
 \hline
$\eta$-percentile &  $\Delta\phi^{st}$ & $N^{st}$ &  $\Delta\phi^{rt}$ & $N^{rt}$ 
  \\ \hline \hline
 $ 0.999$ &  $0.8^{\circ}$ & $166$   &  $1.0^{\circ}$ & $126$
  \\ \hline
 $ 0.99$ &  $2.5^{\circ}$ & $52$   &  $3.2^{\circ}$ & $40$
   \\ \hline
 $ 0.95$ &  $5.6^{\circ}$ & $23$   &  $7.3^{\circ}$ & $18$
   \\ \hline
 $ 0.9$ &  $8.1^{\circ}$ & $16$   &  $10.6^{\circ}$ & $12$
   \\ \hline
 $ 0.75$ &  $14.0^{\circ}$ & $10$   &  $18.3^{\circ}$ & $7$
   \\ \hline
 $ 0.5$ &  $24.7^{\circ}$ & $6$   &  $32.3^{\circ}$ & $4$
   \\ \hline
\end{tabular}
\end{table}

Table \ref{asym} shows a case study with a cluster in a typical indoor environment. The total power, $P_{tot}$ in dBm is calculated via RT-ICM to be $-29.09$. Scan angle is, $\phi_0=53^{\circ}$ and SAS is $S_{\phi}=72.2^{\circ}$. Corresponding Gaussian fitting parameters for RT-ICM are $u=6.43 \times 10^{-5}$ and $v=9.23$ (or $\sigma=9.23/\sqrt{2}=6.52$). Then, from Eq. (\ref{Pmax_conf}) and (\ref{Pmax_icm}), and converting to dBm, $P_{max}^{st}=-19.15$ and $P_{max}^{rt}=-20.87$ both in dBm. In Table \ref{asym}, we calculate $\Delta\phi^{st}$ and $\Delta\phi^{rt}$ numerically using Eq. (\ref{deltaerf_conference}) and (\ref{deltaphi_icm}), respectively. Associated required number of elements to create the beamwidths, $N^{st}$ and $N^{rt}$ are computed using Eq. (\ref{beamwidth_eq}). While infinite number of antenna elements are required to achieve asymptotic values of the received power, Table \ref{asym} shows that high percentile powers can be obtained via practical values of $N$. On the other hand, it can be seen that for $\eta \geq 0.95$, in spite of the beamwidth difference between 802.11ad and RT-ICM is subtle, the difference in the required number of elements is increasing significantly. That proves that the hardware complexity to achieve optimum beamwidth value is directly related (and sensitive) to the correct estimation of the cluster parameters such as angle spread. 

%Finally, it is worthy to note that 802.11ad cluster model is based on the measurements that measure the angle spread \textit{after} processing the signal which basically cuts out the part of the signal that is below the receiver sensitivity. That results in to slightly lower total cluster power and smaller angle spread statistics. On the other hand, beamforming is performed in PHY \textit{before} processing the signal. Thanks to the ray-tracing physics nature, the RT-ICM addresses these problems and provides cluster power and the angle spread in front of the antenna. 

\subsection{Triangular Beam Model} \label{triangular}
Throughout the paper, we used the rectangular beam model given in Eq. (\ref{WR}) so far for all the analysis. In this section, we update the main equations for triangular beam model given in Eq. (\ref{WT}) and provide the difference with respect to the rectangular beam model. 
\paragraph*{Remark 3} Following statements are true for both perfect and imperfect alignments: (1) $\Delta\phi_{opt}$ doesn't depend on the main lobe beam model.  (2) The difference in maximum achievable power is $1.23$ dB.

Starting from the perfect alignment, Eq. (\ref{p}) is calculated as following:
\begin{align*}
\rho&=\int_{\phi_{cl}-\Delta\phi/2}^{\phi_{cl}+\Delta\phi/2} W_T f(\phi) \text{d}\phi \\ 
&= \text{erf} \left( \frac{\Delta\phi}{2\sqrt{2}\sigma} \right) - \frac{\sqrt{2}(1-e^{-\Delta\phi^2/8\sigma^2})}{\sqrt{\pi}\Delta\phi}\numberthis \label{p_trian}
\end{align*}

Plugging into Eq. (\ref{Pextst}), skipping intermediate steps, received power is obtained as
\begin{equation}
P_R(\Delta\phi)=\frac{101.5P_{tot}}{\Delta\phi \sin \phi_0} \text{erf} \left( \frac{\Delta\phi}{2\sqrt{2}\sigma} \right) - \frac{81P_{tot}\sigma (1-e^{-\Delta\phi^2/8\sigma^2})}{\sin \phi_0 \Delta\phi^2} 
\end{equation}

Note that the first term is the $P_R(\Delta\phi)$ for the rectangular beam model given in Eq. (\ref{PR_st_perfect}) which is maximum at $\Delta\phi=0$.  On the other hand, the second term is always positive and minimized at $\Delta\phi=0$. This proves that the $P_{max}$ is at $\Delta\phi=0$. Taking the limit while $\Delta\phi \to 0$, theoretical maximum power with the triangular model is
%\label{Pmax_st}
\begin{equation}
 P_{T,max}^t=P_R(0)=\frac{30.5P_{tot}}{\sigma \sin \phi_0} 
 \end{equation}
 
Comparing with the rectangular model in Eq. (\ref{Pmax_st}), the difference is
\begin{equation}
10 \log \left( P_{max}^t/P_{T,max}^t \right) = 1.23 \ \text{dB}
\end{equation}

Since the result doesn't depend on any parameter, it is also true for the RT-ICM case. 

The proof for the imperfect alignment is performed via simulation as the difference equation is computed numerically due to the \textit{erf}. 

\section{Performance Evaluation} \label{performanceevaluation}
In this section, we give the plots that illustrate the performance of the given analytical expressions by comparing with the simulations. We use the 802.11ad Gaussian notation for the performance results, but, to show the difference between fixed versus exact $\sigma$, we refer 802.11ad for fixed, RT-ICM for the true value. As both models agree for indoor mmWave clusters, we scan a range of $3^{\circ}$ to $10^{\circ}$ for $\sigma$. 

\begin{figure}[t]
\centering
\includegraphics[scale=0.24]{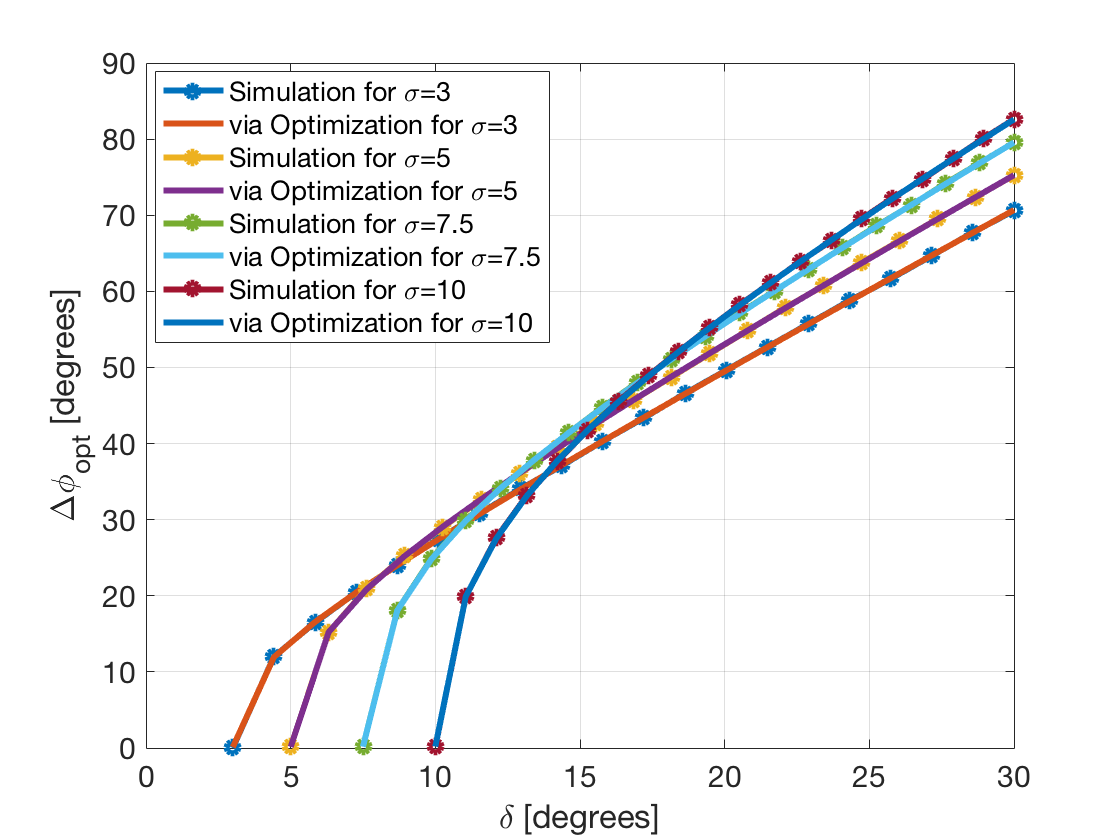} \\
\caption{Accuracy of second derivative test method in case of misalignment and $\delta-\sigma$ relation effect on $\Delta\phi_{opt}$.}\label{simvsopt}
\end{figure}

In the first plot, we measure the accuracy of the method that uses the Second Derivative Test when finding the $\Delta\phi_{opt}$ in case of misalignment, given in Sec. \ref{PR_maximization_imperfect}. As seen from Fig. \ref{simvsopt}, optimization results for $\Delta\phi_{opt}$ perfectly match to simulation results for different values of $\sigma$ and $\delta$, as stated in \textit{Remark 1}. On the other hand, note that $\Delta\phi_{opt} \to 0$ rapidly while $\delta \approx \sigma$. Also, it can be seen that $\Delta\phi_{opt} > 2\delta$ once  $\delta \geq \sqrt{2}\sigma$. This concludes the proof of  \textit{Remark 2}.

%\begin{figure}[t]
%\centering
%\includegraphics[scale=0.24]{deltavsbeam.png} \\
%\caption{$\Delta\phi_{opt}$ with changing $\delta$ and $\sigma$ and $\delta-\sigma$ relation.}\label{deltavsbeam}
%\end{figure}

\begin{figure}[t]
\centering
\includegraphics[scale=0.24]{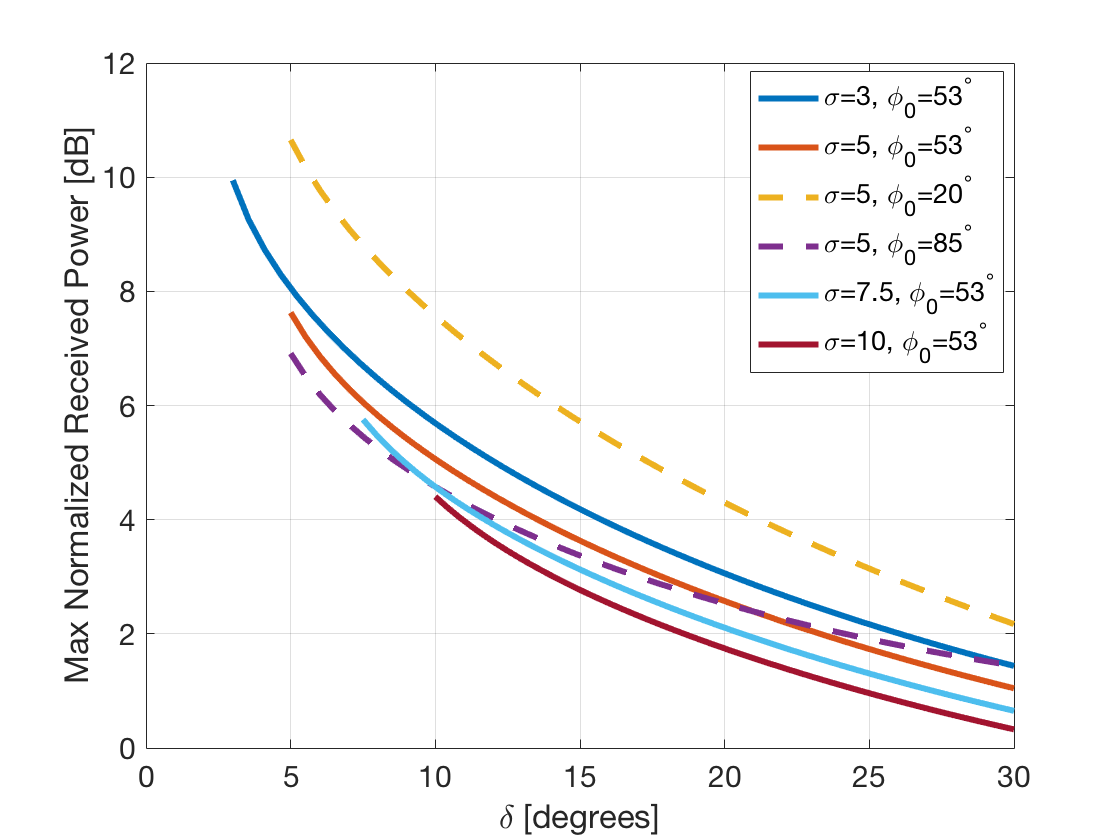} \\
\caption{Variation on the maximum achievable received power with respect to $\delta$ for different $\sigma$ and $\phi_0$.}\label{Pmax_normalized}
\end{figure}

Fig. \ref{Pmax_normalized} shows how the maximum received power given in Eq. (\ref{Pmaximperfect}) changes in dB scale with variation in $\delta$ for four different $\sigma$s. In the figure, the received power is normalized with $P_{tot}$. As seen, once $\delta > \sigma$, $P_{max}$ drops exponentially and becomes the dominant degradation factor. However, even with a large $\delta$, using a ULA has still advantages over an omnidirectional antenna. On the other hand, as $\delta \approx \sigma$, $P_{max}$ reaches to its max level. Although not shown here, for $\delta < \sigma$, it saturates rapidly, which can be concluded from Fig. \ref{PRimperfect}. That results in the following important conclusion: As long as $\delta < \sigma$, the misalignment error is tolerable thanks to the antenna gain and maximum power can be still achieved. 

%It is worthy to stress here that we ignore the sidelobes in this analysis which may add a little extra power.

Fig. \ref{Pmax_normalized} also shows the effect of the steering angle $\phi_0$ to $P_{max}$ for $\sigma=5$. As seen, $P_{max}$ strictly depends on $\phi_0$ due to the $sin$ factor in the denominator. However, recall that since $\phi_0$ depends on the channel parameter $\phi_{cl}$, user has limited control over it. 

%For completeness, we will use the same case study defined in the previous section. 

\begin{figure}[t]
\centering
\includegraphics[scale=0.24]{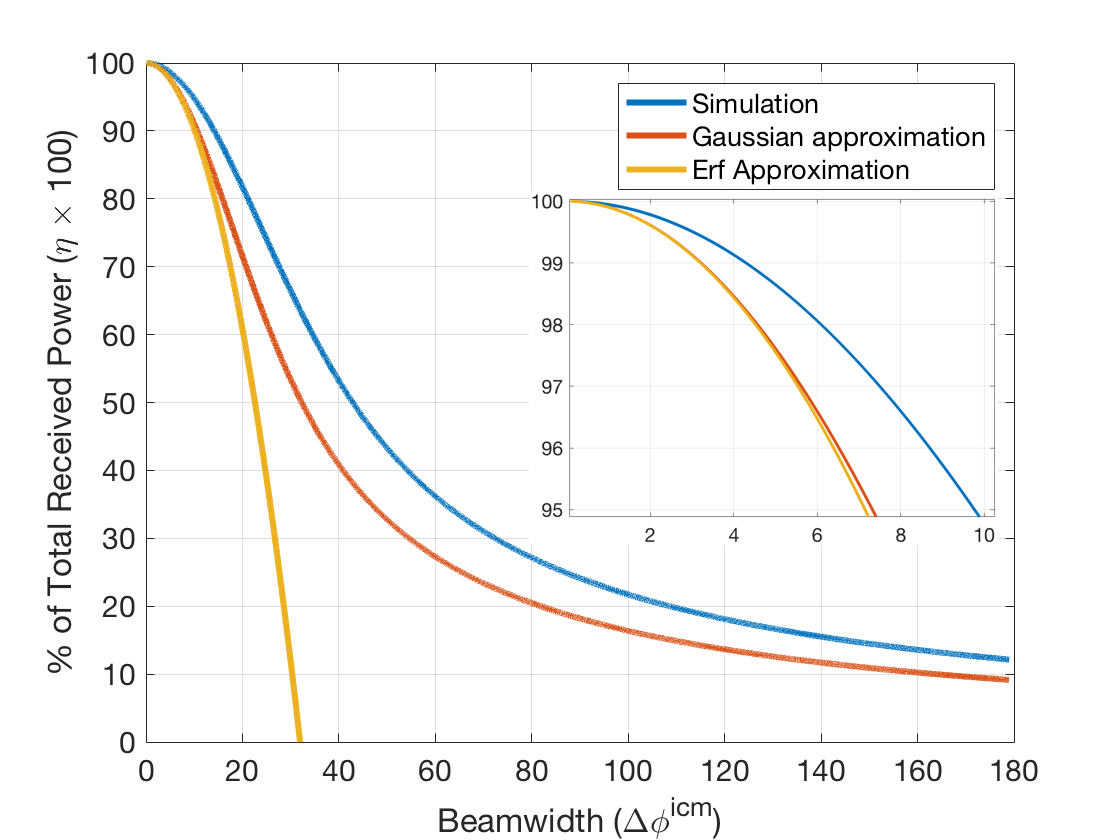} \\
\caption{Practical beamwidth analysis performance of RT-ICM for $v=9.23$ (or $\sigma=6.52$).}\label{perform_icm}
\end{figure}

To illustrate the performance of the analysis in the case of perfect alignment (or $\delta<\sigma$), in Fig. \ref{perform_icm}, we demonstrate the performance of RT-ICM expressions given in Eq. (\ref{deltaphi_icm}) and (\ref{deltaphi_icm_approx}) with respect to the percentile power in percentage, for $v=9.23$. We use Eq. (\ref{PextICM}) to simulate the model without any approximation, shown with a blue line in the figure. The red curve is the resulted relation of Gaussian approximation given in Eq. (\ref{deltaphi_icm}), also some values are given in Table \ref{asym}. Finally, the orange line is for the error function approximation given in Eq. (\ref{deltaphi_icm_approx}) for the given condition that $\Delta\phi \leq 2v=18.46$.  Note that the approximation of the $erf$ is in an almost perfect agreement with its numeric solution within the supported range. On the other hand, Gaussian fitting yields a result with a little error compared to simulation. However, for example, for $\eta=0.99$, the beamwidth error is only $1^{\circ}$.

%\begin{figure}[t]
%\centering
%\includegraphics[scale=0.24]{perform_st.png} \\
%\caption{Practical beamwidth analysis performance of IEEE 802.11ad model.}\label{perform_st}
%\end{figure}

%Similar performance comparison for the IEEE 802.11ad model beamwidth expressions derived in Sec. \ref{problem_st} are plotted in Fig. \ref{perform_st}. In the figure, numeric solution of Eq. (\ref{deltaerf_conference}) (shown with red line) fits simulation results almost perfectly whereas the approximate solution to beamwidth in Eq. (\ref{beamwidth_11ay_conf}) (plotted with orange line) drifts away once $\Delta\phi \gtrapprox 15^{\circ}$ which supports the discussion made during the equation derivation. However, as also seen in the zoomed figure, for $\eta > 0.95$, Eq. (\ref{beamwidth_11ay_conf}) estimates the optimum beamwidth with a high accuracy. For example, the beamwidth error caused by the approximation is only $0.2^{\circ}$ for $\eta=0.99$.

\begin{figure}[t]
\centering
\includegraphics[scale=0.24]{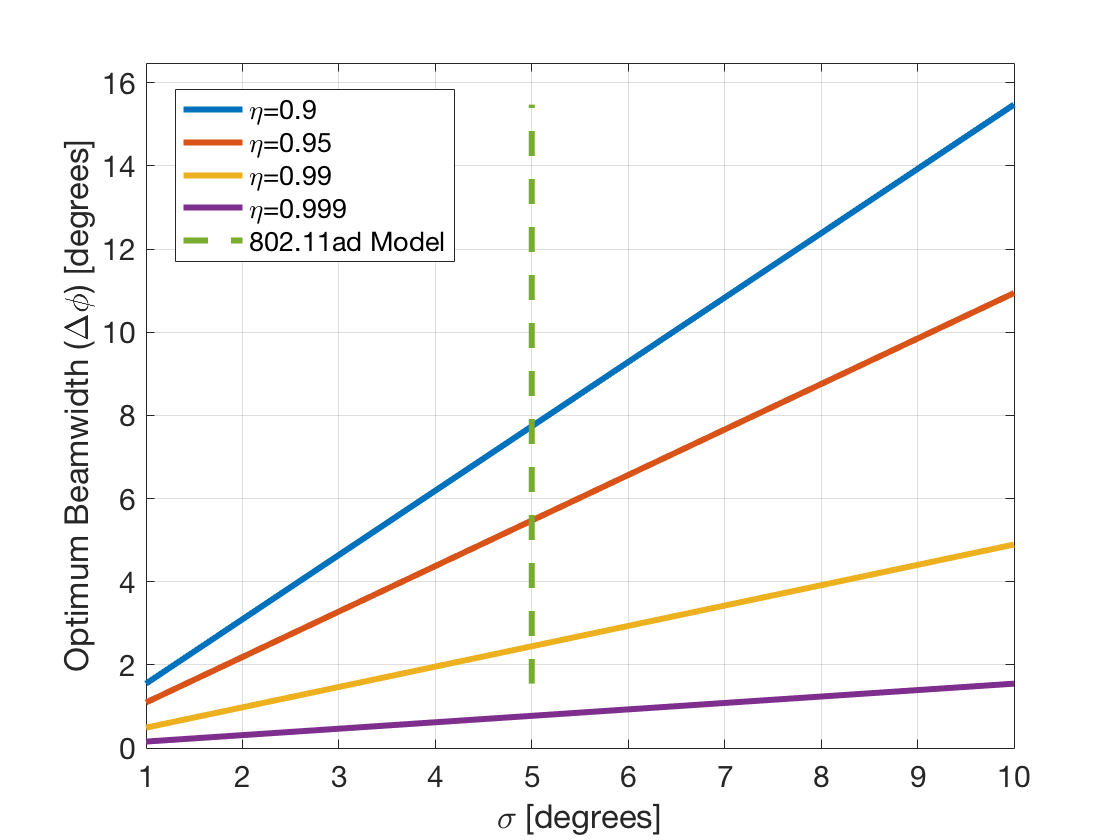} \\
\caption{Beamwidth Error in IEEE 802.11ad model due to fixed $\sigma=5^{\circ}$ in the perfect alignment case.}\label{sigmadrifts}
\end{figure}

%Note from the relation between the 802.11ad and RT-ICM approximate beamwidth solutions in Eq. (\ref{beamwidth_11ad}) and  (\ref{deltaphi_icm_approx}), we conclude $v=\sqrt{2}\sigma$ which can also be seen from the Gaussian function definition given in Eq. (\ref{gphi}). 
In Fig. \ref{sigmadrifts}, beamwidth error due to the fixed $\sigma=5^{\circ}$ assumption in IEEE 802.11ad model is plotted for different percentile values. For ease of illustration, $\sigma=5^{\circ}$ line is also stressed. It is seen that the optimum beamwidth sensitivity to $\sigma$ increases if a low $\eta$ is selected. For interference-limited systems where high power and directed beams are needed, the 802.11ad model works with subtle beamwidth errors, whereas for hardware-limited systems with a limited number of antenna elements, the 802.11ad model may estimate the optimum beamwidth with significant errors. Also, it is worthy to note that the increase (or decrease) rate at the required number of antenna elements ($N$) in case of an error is the same for all $\eta$ values. 

\begin{figure}[t]
\centering
\includegraphics[scale=0.24]{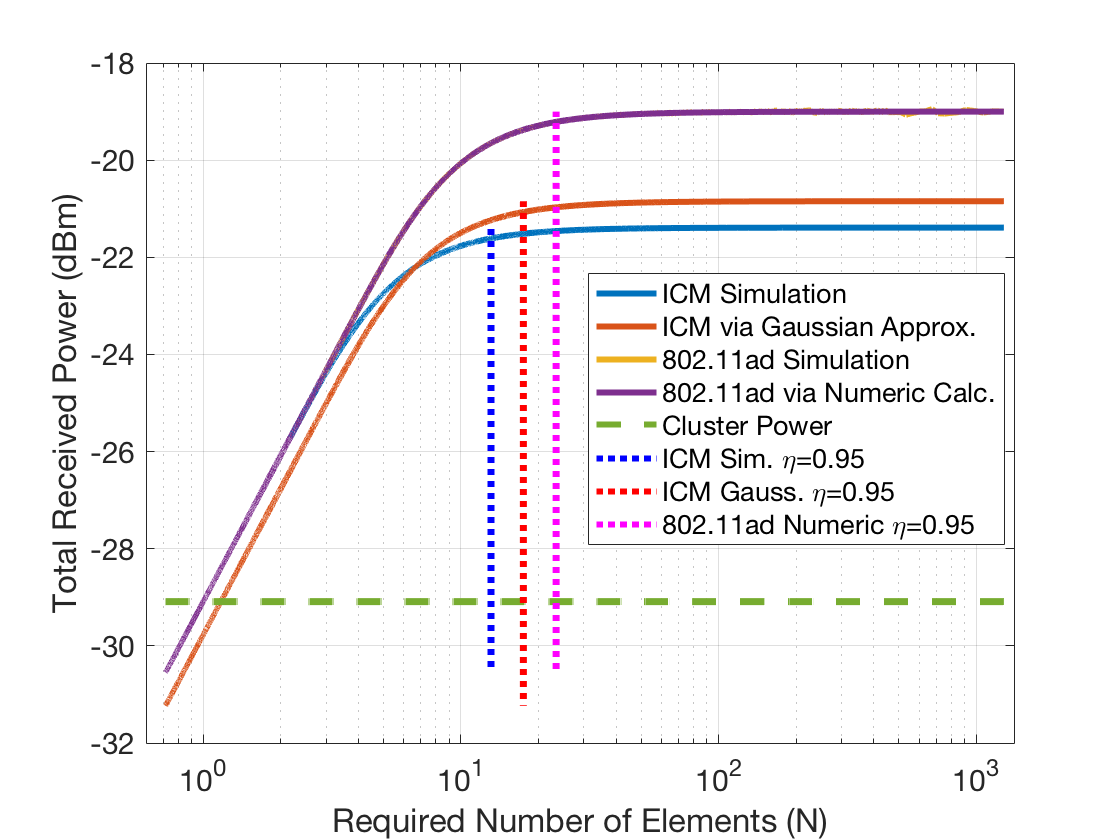} \\
\caption{Models performance comparison in terms of required antenna elements for perfect alignment}\label{antelemcomparison}
\end{figure}

Fig. \ref{antelemcomparison} demonstrates the total received power performance of both models with respect to the number of antenna elements that is required to create the optimum beamwidth. We plot the simulation for two models along with their corresponding received power equations given in (\ref{PR_st_perfect}) and (\ref{PRdelta_icm}), respectively. The same case study, given in Sec. \ref{asymptotic_perfect}, is used here. For the abscissa, we use the Eq. (\ref{beamwidth_eq}) for the transformation from $\Delta\phi$ to $N$. In the figure, $\eta=0.95$ percentile points are also denoted with dotted lines. Recalling the high accuracy of the RT-ICM, we accept the RT-ICM simulation as the most accurate curve and compare the other realizations with it. As seen from the figure, 802.11ad model yields $\sim2$ dB more total power both with simulation and analytical expression, for this case study. This effect is due to the smaller angle spread measurements of 802.11ad. It can also be confirmed from Eq. (\ref{Pmax_st}) where angle spread standard deviation is inverse relation with maximum received power. While $\sigma=5$, fitting Gaussian function of RT-ICM angle spread standard deviation is estimated to be $v/\sqrt{2}=6.52$ for this case study. 

On the other hand, Fig. \ref{antelemcomparison} shows that the required number of elements ($N$) to create a beamwidth that would achieve $\%95$ of the maximum total power is $13$, as shown with the blue dotted line. The Gaussian approximation of RT-ICM yields $N=18$ whereas the 802.11ad model requires $23$ elements to reach its $\%95$ of maximum received power.
%That should be mainly because of the \textit{after} processing angle spread measurement of the 802.11ad which results in lower angle spread. Since we set equal total cluster powers for both ICM and 802.11ad, captured power within a certain beamwidth for 802.11ad is larger than that of ICM. 

\begin{figure}[t]
\centering
\includegraphics[scale=0.24]{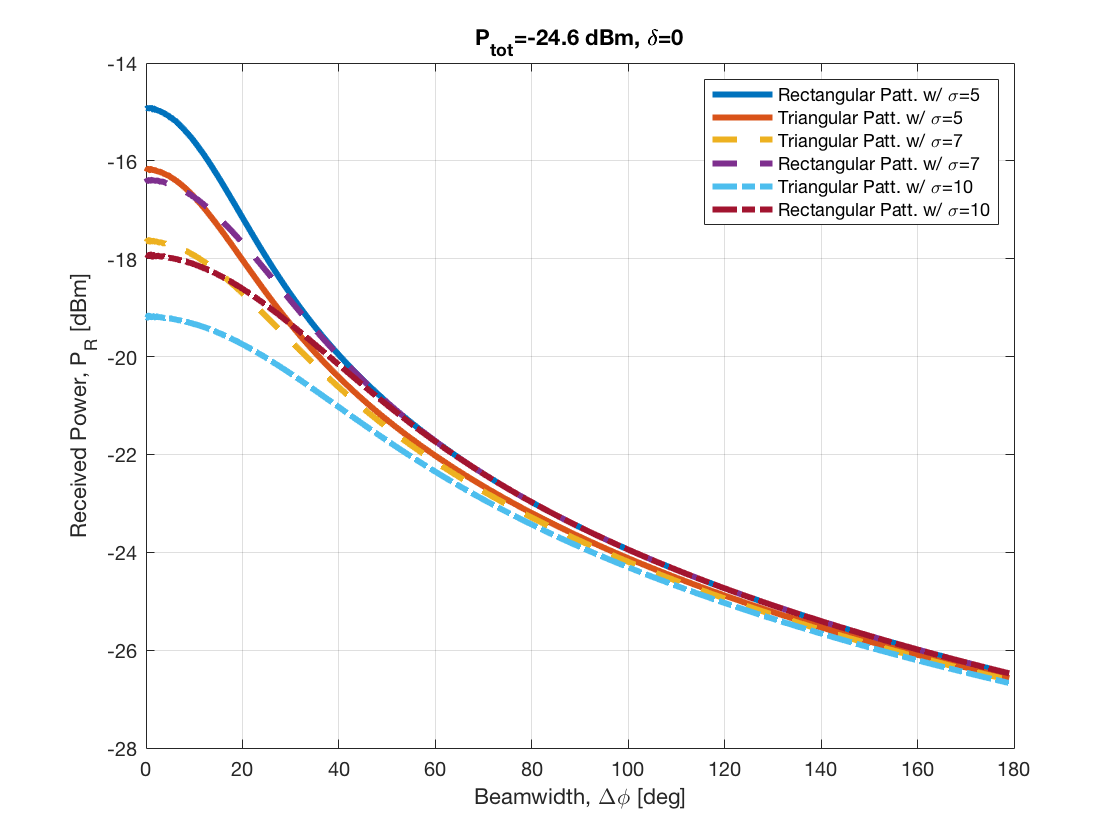} \\
\caption{Beamwidth Error in IEEE 802.11ad model due to fixed $\sigma=5^{\circ}$ in the perfect alignment case.}\label{tri_vs_rect}
\end{figure}

\begin{figure}[t]
\centering
\includegraphics[scale=0.24]{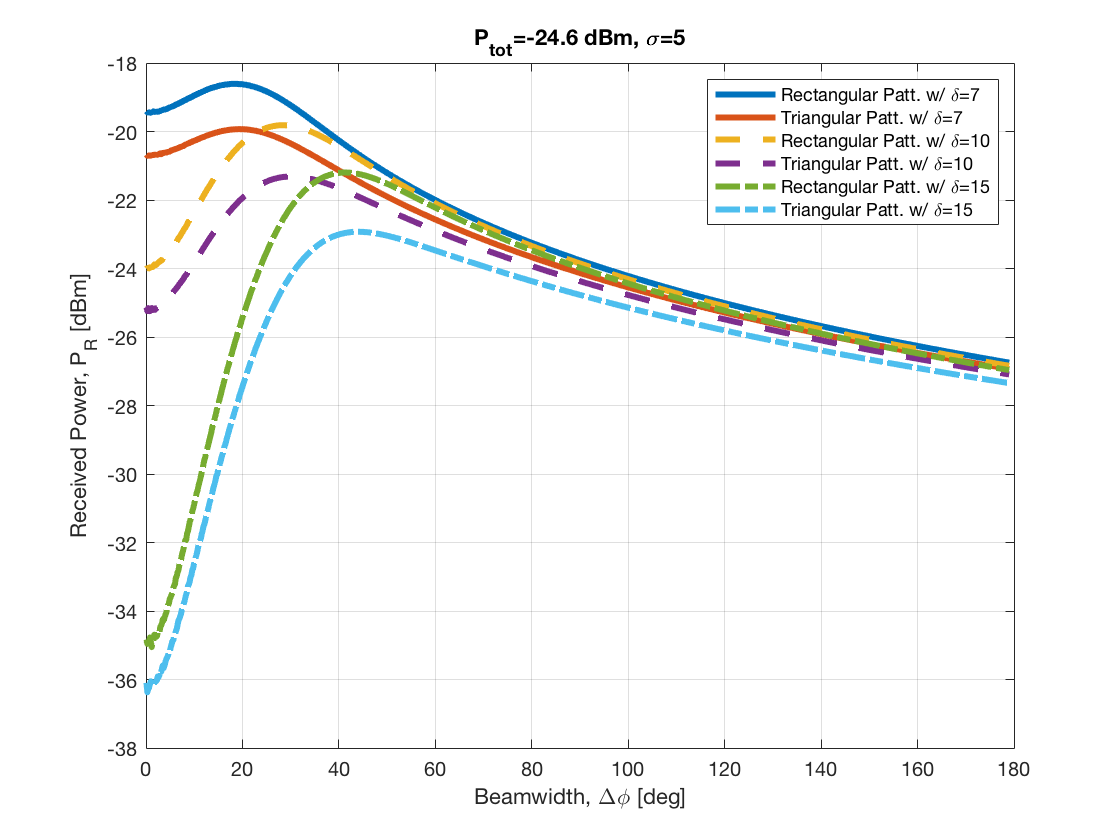} \\
\caption{Models performance comparison in terms of required antenna elements for perfect alignment}\label{trianglewhenmisal}
\end{figure}

Finally, Fig. \ref{tri_vs_rect} and \ref{trianglewhenmisal} show the triangular and the rectangular beam model comparison, for perfect and imperfect alignment, respectively. As derived in Sec. \ref{triangular}, while $\Delta\phi_{opt}$ is unchanged, $P_{max}$ reduces $1.23$ dB for the triangular model. Note the true shape of the main lobe is more like Gaussian, and its $P_{max}$ should reside between rectangular and triangular model as shown in Fig. \ref{patternapprox}. Hence, using the rectangular model for the main lobe doesn't create significant errors while complexity in the derivation gets extremely simple.  

\section{Conclusion} \label{conclusion}
In this paper, we provide an analytical framework for the optimum beamwidth that maximizes the received power for indoor mmWave clusters. In the analysis, we consider the uniform linear array (ULA) antenna type for the antenna gain, and two beam models to approximate the main lobe array pattern; rectangular and triangular. Expressions that relate the beamwidth and captured cluster channel gain is provided for two intra-cluster model, IEEE 802.11ad and Ray Tracing based Intra-Cluster Model (RT-ICM) and the optimization problem is introduced by combining the antenna gain. Both perfect and imperfect alignment scenarios are studied. For misalignment cases, the optimum beamwidth is found to be larger than the 2 times of the alignment error when the error is larger than the $\sqrt{2}$ of the standard deviation of the cluster channel.   For perfect alignment, we show that the theoretical maximum received power converges to a constant while optimum beamwidth approaches to zero. We then provide equations that would result in practical beamwidth values while sacrificing from the maximum received power in the order of tenths percentage. Finally, we evaluate the performance of the analysis by comparing the analytical results with simulations for an indoor mmWave cluster. The work we propose in this paper will give insights to the optimum antenna array design in both MIMO and massive MIMO applications for future mmWave systems. 

\appendices
\section{Maximization of $P_R$ in case of Misalignment} \label{maxPRmisalign}
\subsection{First Derivative}
First derivative of Eq. (\ref{maxPR}) and equalize zero,
\begin{align*}
&\frac{d \left( \frac{1}{\Delta\phi} \left( \text{erf} \left( \frac{\Delta\phi + 2\delta}{2\sqrt{2} \sigma} \right) +  \text{erf} \left( \frac{\Delta\phi - 2\delta}{2\sqrt{2} \sigma} \right) \right) \right) }{d \Delta\phi} \\
&= \left( \frac{\text{erf} \left( \frac{\Delta\phi + 2\delta}{2\sqrt{2} \sigma} \right)}{\Delta\phi} \right)' +  \left( \frac{\text{erf} \left( \frac{\Delta\phi - 2\delta}{2\sqrt{2} \sigma} \right)}{\Delta\phi} \right)' =0
\end{align*}

where 
\begin{equation*}
\left( \frac{\text{erf} \left( \frac{\Delta\phi + 2\delta}{2\sqrt{2} \sigma} \right)}{\Delta\phi} \right)'=\frac{\frac{e^{-(\Delta\phi+2\delta)^2/(8\sigma^2)}}{\sqrt{2\pi}\sigma}\Delta\phi - \text{erf} \left( \frac{\Delta\phi + 2\delta}{2\sqrt{2}\sigma} \right)}{(\Delta\phi)^2}
\end{equation*}
\begin{equation*}
\left( \frac{\text{erf} \left( \frac{\Delta\phi - 2\delta}{2\sqrt{2} \sigma} \right)}{\Delta\phi} \right)'=\frac{\frac{e^{-(\Delta\phi-2\delta)^2/(8\sigma^2)}}{\sqrt{2\pi}\sigma}\Delta\phi - \text{erf} \left( \frac{\Delta\phi - 2\delta}{2\sqrt{2}\sigma} \right)}{(\Delta\phi)^2}
\end{equation*}
Summing up, Eq. (\ref{eqn_dbl_x}) is obtained.
\begin{figure*}[!t]
% ensure that we have normalsize text
\normalsize
% Store the current equation number.
\setcounter{mytempeqncnt2}{\value{equation}}
% Set the equation number to one less than the one
% desired for the first equation here.
% The value here will have to changed if equations
% are added or removed prior to the place these
% equations are referenced in the main text.
%\setcounter{equation}{44}
\begin{equation}
\label{eqn_dbl_appen}
\frac{d^2 \left( \frac{1}{\Delta\phi} \left( \text{erf} \left( \frac{\Delta\phi + 2\delta}{2\sqrt{2} \sigma} \right) +  \text{erf} \left( \frac{\Delta\phi - 2\delta}{2\sqrt{2} \sigma} \right) \right) \right) }{d^2 \Delta\phi} =\left( \frac{ e^{\frac{-(\Delta\phi+2\delta)^2}{8\sigma^2}} + e^{\frac{-(\Delta\phi-2\delta)^2}{8\sigma^2}}}{\sqrt{2\pi}\sigma \Delta\phi} \right)'  - \left( \frac{\text{erf} \left( \frac{\Delta\phi + 2\delta}{2\sqrt{2}\sigma} \right) + \text{erf} \left( \frac{\Delta\phi - 2\delta}{2\sqrt{2}\sigma} \right)}{(\Delta\phi)^2} \right)'<0
\end{equation}
\begin{equation}
\label{eqn_dbl_appen2}
\left( \frac{ e^{\frac{-(\Delta\phi+2\delta)^2}{8\sigma^2}} + e^{\frac{-(\Delta\phi-2\delta)^2}{8\sigma^2}}}{\sqrt{2\pi}\sigma \Delta\phi} \right)' = \frac{-\Delta\phi(\Delta\phi(\Delta\phi+2\delta)+4\sigma^2)e^{\frac{-(\Delta\phi+2\delta)^2}{8\sigma^2}}-\Delta\phi(\Delta\phi(\Delta\phi-2\delta)+4\sigma^2)e^{\frac{-(\Delta\phi-2\delta)^2}{8\sigma^2}}}{4\sigma^3\sqrt{2} (\Delta\phi)^2}
\end{equation}
\begin{equation}
\label{eqn_dbl_appen3}
\left( \frac{\text{erf} \left( \frac{\Delta\phi + 2\delta}{2\sqrt{2}\sigma} \right) + \text{erf} \left( \frac{\Delta\phi - 2\delta}{2\sqrt{2}\sigma} \right)}{(\Delta\phi)^2} \right)'= \frac{\Delta\phi \left( e^{\frac{-(\Delta\phi+2\delta)^2}{8\sigma^2}} + e^{\frac{-(\Delta\phi-2\delta)^2}{8\sigma^2}} \right) -2\sqrt{2\pi}\sigma  \left( \text{erf} \left( \frac{\Delta\phi + 2\delta}{2\sqrt{2}\sigma} \right) + \text{erf} \left( \frac{\Delta\phi - 2\delta}{2\sqrt{2}\sigma} \right)  \right)}{\sigma \sqrt{2\pi} (\Delta\phi)^3}
\end{equation}
% Restore the current equation number.
%\setcounter{equation}{\value{mytempeqncnt2}}
% IEEE uses as a separator
\hrulefill
% The spacer can be tweaked to stop underfull vboxes.
\vspace*{2pt}
\end{figure*}

\subsection{Second Derivative}
Second derivative of Eq. (\ref{maxPR}) with setting smaller than zero is given in Eq. (\ref{eqn_dbl_appen})
through (\ref{eqn_dbl_appen3}). After subtraction and modification such that the denominator is $4\sigma^3 \sqrt{2\pi} (\Delta\phi)^3$, the denominator can be eliminated as it is positive and equation results in Eq. (\ref{eqn_dbl_y}).

\section{Maximization of $P_R$ for Perfect Alignment} \label{802.11ad}
\subsection{Maximum Received Power Derivation}
Derivatives of the nominator and denominator of Eq. (\ref{PR_st_perfect}) are,
\begin{equation*}
\frac{d \left( \text{erf} \left( \frac{\Delta\phi}{2\sqrt{2}\sigma} \right) \right)}{d \Delta\phi}=\frac{e^{-\Delta\phi^2/8\sigma^2}}{\sigma \sqrt{2\pi}}
\end{equation*}
\begin{equation*}
\frac{d \left( \frac{\Delta\phi \sin \phi_0}{101.5P_{tot}} \right)}{d \Delta\phi} =\frac{\sin \phi_0}{101.5P_{tot}}
\end{equation*}
 
Then, the maximum received power is 
 \begin{align*}
 P_{max}^{t}=\lim_{\Delta\phi \to 0} P_R(\Delta\phi)&=\frac{101.5P_{tot}}{\sin \phi_0} \frac{e^{-\Delta\phi^2/8\sigma^2}}{\sigma \sqrt{2\pi}} |_{\Delta\phi=0}\\
 &=\frac{40.5P_{tot}}{\sigma \sin \phi_0}
 \end{align*}

\subsection{Optimum Beamwidth Approximation}

Using the first and second term of the error function Taylor expansion, Eq. (\ref{deltaerf}) becomes
\begin{equation*}
\frac{\Delta\phi_{\eta}}{\left( \frac{\Delta\phi_{\eta}}{2\sqrt{2}\sigma}- \frac{(\Delta\phi_{\eta}/2\sqrt{2}\sigma)^3}{3} \right)}  =  \frac{5\sigma}{\eta\sqrt{\pi}}
\end{equation*}

Further manipulation gives
\begin{equation*}
(\Delta\phi_{\eta})^3-(24\sigma^2)\Delta\phi_{\eta}  + \eta  (24.06 \sigma^2)\Delta\phi_{\eta} =0
\end{equation*}

Ignoring the hundredths decimal and solving the equation for $\Delta\phi_{\eta}>0$, Eq. (\ref{beamwidth_11ad}) results.

%\section{RT-ICM Derivations} \label{ICM_appen}
%\subsection{Maximum Power Derivation}
%Analogous to the previous subsection, maximum received power is achieved while $\Delta\phi \to 0$ and the limit has the $0/0$ indeterminate form. Taking the derivative of the numerator and the denominator,
%\begin{equation*}
%\frac{d \left( \text{erf} \left( \frac{\Delta\phi}{2v} \right) \right)}{d \Delta\phi}=\frac{e^{-\Delta\phi^2/4v^2}}{v \sqrt{\pi}}
%\end{equation*}
%\begin{equation*}
%\frac{d \left( \frac{\Delta\phi \sin \phi_0}{101.5uv\sqrt{\pi}} \right)}{d \Delta\phi} =\frac{\sin \phi_0}{101.5uv\sqrt{\pi}}
%\end{equation*}

%Then, the maximum received power for the RT-ICM is
 %\begin{equation}
% P_{max}^{rt}=\frac{101.5uv\sqrt{\pi}}{\sin \phi_0} \frac{e^{-\Delta\phi^2/4v^2}}{v \sqrt{\pi}} |_{\Delta\phi=0} 
 %\end{equation}
%\subsection{Optimum Beamwidth Approximation}
%Similar to 802.11ad case, using the first two terms of Taylor expansion,
%\begin{equation*}
%\frac{\Delta\phi}{\left( \frac{\Delta\phi}{2v}- \frac{(\Delta\phi/2v)^3}{3} \right)} =\frac{2v}{\eta }
%\end{equation*}

%Solving for $\Delta\phi=\Delta\phi^{icm}$ gives Eq. (\ref{deltaphi_icm_approx}).

\section*{Acknowledgment}
The authors would like to thank Prof. S. Orfanidis for many helpful discussions and his contributions to the antenna theory. 

\bibliographystyle{IEEEtran}

%Authors
%\newpage
\begin{IEEEbiographynophoto}{Yavuz Yaman}
received the B.S degree from the School of Engineering, Istanbul University, in 2011; M.S. degree in electrical and computer engineering from Rutgers University, Piscataway, NJ, in 2014. He is currently working toward the Ph.D. degree with the Department of Electrical and Computer Engineering, Rutgers University, Piscataway, NJ. His research interests include channel modelling, beamforming, channel estimation, antenna propagations and phased antenna arrays.
\end{IEEEbiographynophoto}

\begin{IEEEbiographynophoto}{Predrag Spasojevic}
received the Diploma of Engineering degree from
the School of Electrical Engineering, University of Sarajevo, in 1990;
and M.S. and Ph.D. degrees in electrical engineering from Texas A\&M
University, College Station, Texas, in 1992 and 1999, respectively.
From 2000 to 2001, he was with WINLAB, Electrical and Computer
Engineering Department, Rutgers University, Piscataway, NJ, as a
Lucent Postdoctoral Fellow. He is currently  Associate Professor in
the Department of Electrical and Computer Engineering, Rutgers
University. Since 2001 he is a member of the WINLAB research faculty.
 From 2017 to 2018 he was a Senior Research Fellow with
Oak Ridge Associated Universities working at the Army Research Lab, Adelphi, MD.
 His research interests are in the general areas of
communication and information theory, and signal processing.

Dr. Spasojevic was an Associate Editor of the IEEE Communications
Letters from 2002 to 2004 and served as a co-chair of the DIMACS
Princeton-Rutgers Seminar Series in Information Sciences and Systems
2003-2004. He served as a Technical Program Co-Chair for IEEE Radio
and Wireless Symposium in 2010. From 2008-2011 Predrag served as the
Publications Editor of IEEE Transactions of Information Theory.
\end{IEEEbiographynophoto}

%\newpage


\end{document}